\documentclass[10pt]{iopart}
\usepackage{iopams, setstack}
\usepackage{bm,hyperref}

\newcommand{\eqref}[1]{{(\ref{#1})}}    

\newcommand{\rmSelf}{{\rm self}}    
\newcommand{\rmExt}{{\rm ext}}
\newcommand{\rmS}{{\rm S}}
\newcommand{\rmR}{\mathrm{R}}

\newcommand{\ItP}{{\mathcal{P}}} 
\newcommand{\ItF}{{\mathcal{F}}} 
\newcommand{\ItX}{{\mathcal{X}}}
\newcommand{\ItY}{{\mathcal{Y}}}

\newcommand{\GS}{{G_{\mathrm{S}}}}     
\newcommand{\GR}{{G_{\mathrm{R}}}}      
\newcommand{\phiS}{{\phi_{\mathrm{S}}}}
\newcommand{\phiR}{{\phi_{\mathrm{R}}}}

\newcommand{\Lie}{{\mathcal{L}}} 
\newcommand{\LieX}{{\mathcal{L}_{\xi}}} 

\begin{document}
\title{Effective stress-energy tensors, self-force, and broken symmetry}
\author{Abraham I. Harte}
\address{Enrico Fermi Institute}
\address{University of Chicago, Chicago, IL 60637 USA}
\ead{harte@uchicago.edu}

\date{December 18, 2009}

\begin{abstract}
Deriving the motion of a compact mass or charge can be complicated by the presence of large self-fields. Simplifications are known to arise when these fields are split into two parts in the so-called Detweiler-Whiting decomposition. One component satisfies vacuum field equations, while the other does not. The force and torque exerted by the (often ignored) inhomogeneous ``S-type'' portion is analyzed here for extended scalar charges in curved spacetimes. If the geometry is sufficiently smooth, it is found to introduce effective shifts in all multipole moments of the body's stress-energy tensor.  This greatly expands the validity of statements that the homogeneous R field determines the self-force and self-torque up to renormalization effects. The forces and torques exerted by the S field directly measure the degree to which a spacetime fails to admit Killing vectors inside the body. A number of mathematical results related to the use of generalized Killing fields are therefore derived, and may be of wider interest. As an example of their application, the effective shift in the quadrupole moment of a charge's stress-energy tensor is explicitly computed to lowest nontrivial order.
\end{abstract}


\vskip 2pc

\section{Introduction}

Essentially all observational consequences of general relativity rely in some way on understanding how objects move in response to the curvature of spacetime. This problem has been extensively studied in many contexts. One area in which some uncertainties still remain relates to the ``bulk'' interaction of matter with its own (gravitational or other) long-range fields. Self-force effects such as these were first studied in flat spacetime electromagnetism. At the time, the main motivation was to explain the then newly-discovered electron. Today, interest in the problem has been revived in order to better understand the gravitational self-force involved in the decay of astrophysical compact binaries.

The first work on the electromagnetic self-interaction problem involved direct calculations of the force acting on a charged rigid sphere in flat spacetime \cite{Abraham, Lorentz, Schott}. At least in the nonrelativistic limit, the acceleration $\vec{a}$ of the sphere's center was found to be closely approximated by
\begin{equation}
    m \vec{a} = \vec{F}_\rmExt + \frac{2}{3} q^2 \dot{\vec{a}} - \delta m \vec{a} .
    \label{nonRelRR}
\end{equation}
$q$ represents the body's total charge and  $m$ its mass (as calculated by integrating over the interior in the usual way). $\vec{F}_\rmExt$ is the ordinary force exerted by the external electromagnetic field. The last two terms on the right-hand side of \eqref{nonRelRR} are due to the self-field. The first represents the expected reaction to emitted radiation. The other effectively shifts the charge's inertial mass. The formula for $\delta m$ is relatively complicated. It has no meaningful point particle limit, and is strongly dependent on the details of the charge's internal structure.

For these reasons and others, it has been common to interpret the mass renormalization term as ``practically unobservable'' in most contexts. The inertial mass of a particle is usually inferred by observing its reaction to applied forces. It is clear from the above equation that such a mass is given by $m_{\mathrm{eff}} = m + \delta m$. The relation between $m_\mathrm{eff}$ and the body's internal structure is often irrelevant. It is just some parameter that can be fixed experimentally (at least if there is reason to expect $\dot{m}_{\mathrm{eff}} = 0$).

This point of view has led to methods that directly extract only the more interesting ``radiative'' portion of the self-force. One proposal -- originally due to Dirac \cite{Dirac} -- is to compute the force exerted by one-half of the difference between the charge's retarded and advanced fields. The result satisfies the vacuum Maxwell equations, and almost always varies slowly inside the body. It is also insensitive to the details of the charge's internal structure. A simple calculation of the force that it exerts recovers the classical Abraham-Lorentz-Dirac expression with far less work than most other methods. Similar ideas have recently been generalized by Detweiler and Whiting in order to recover standard results for scalar, electromagnetic, and gravitational self-forces in curved backgrounds \cite{DetWhiting}. In any of its incarnations, the effects of self-interaction on a body's motion can be interpreted as arising only from a particular vacuum component of its self-field (up to ``renormalization effects'').

We will refer to this idea as the Detweiler-Whiting axiom. It was originally applied to point particles as a physical explanation and a shortcut to the conclusions of various perturbative calculations. It was not, however, derived in any rigorous sense. Its accuracy and range of validity were therefore unknown. This situation has recently changed. A direct derivation of a suitably interpreted Detweiler-Whiting axiom is now available in very general contexts. Natural linear and angular momenta were non-perturbatively defined for nearly arbitrary extended bodies in curved spacetimes. These were shown to have evolution equations that depended only weakly on the inhomogeneous ``S-type'' self-field \cite{HarteScalar, HarteEMNew, HarteGrav}. The main effect of this field was absorbed into the definitions of the momenta; a direct generalization of the mass renormalization effect illustrated by \eqref{nonRelRR}. It is the only consequence of the S field in Minkowski and de Sitter backgrounds. More generally, the existence of any single Killing vector implies that the evolution equation for a particular component of the momentum is completely independent of the S field. If there are no background symmetries, a similar result can still be derived to lowest order in a standard approximation scheme. The S component of the self-field has a small effect in this case, however. It does more than shift the effective momenta.

This paper discusses the nature of these extra forces. Mathematically, they arise as a measurement of the degree to which a particular Green function fails to be invariant under the action of certain generalized Killing fields. The result of this broken symmetry is shown to have a very simple interpretation. The quadrupole and higher multipole moments of a body's stress-energy tensor are effectively shifted by small amounts. In combination with the aforementioned momentum renormalization, this demonstrates that the only direct effect of the S field is to make an object move as though it had a different stress-energy tensor. The effective stress-energy tensor couples to the gravitational tidal fields, just as it would even for a test body in Newtonian theory. The self-field therefore influences a body's ``mass distribution'' in both inertial and (at least passively) gravitational senses.

The framework used to derive these results has been developed for objects individually coupled to scalar, electromagnetic, or (linearized) gravitational fields \cite{HarteScalar, HarteEMNew, HarteGrav}. A set of generalized Killing vectors must be defined first \cite{HarteSyms}. This requires a preferred worldline and a set of spacelike hypersurfaces foliating the object's worldtube. Both of these constructions are usually fixed using center of mass conditions. Linear and angular momenta for general compact matter distributions are then given as conjugate to each generalized Killing field. Their evolution equations are determined using stress-energy conservation. If the homogeneous R-type self-field is not too large, the laws of motion in this formalism coincide with those derived by Dixon \cite{Dix70a,Dix74,Dix79} in the test body limit.

These ideas are not widely known, so a review of the theory as it applies to scalar fields is presented in Sect. \ref{Sect: Review}. Those results are then used in Sect. \ref{Sect: EffMoments} to argue that a charge's S field only acts to (finitely) renormalize the multipole moments of its stress-energy tensor. Sect. \ref{Sect: ExpandLie} develops several mathematical results needed to use generalized Killing fields in perturbation theory. Some results related to bitensors, Riemann normal coordinates, and metric normal tensors used there are collected in the appendix. As a simple application of these ideas, the shift in the effective quadrupole moment of scalar charge is finally computed in Sect. \ref{Sect: Quadrupole}.

The metric is chosen to have signature $-1,1,1,1$. Units are used such that $G=c=1$, and sign conventions for the curvature follow those of Wald \cite{Wald}. Lower case Latin letters $a,b,\ldots$ are used to denote abstract indices. Upper case Latin letters are used as coordinate indices. $A,B,\ldots$ run from $0$ to $3$, while $I,J,\ldots$ can take the values $1$ to $3$.

\section{Mechanics in curved spacetime}
\label{Sect: Review}

Useful notions of linear and angular momenta are not easily defined for extended bodies in generic curved spacetimes. Various concepts have been advanced over the years in order to solve different problems. A particularly interesting example was proposed by Dixon in the 1970's \cite{Dix70a, Dix74, Dix79, Dix67, BaileyIsrael}. His momenta arose as part of a general theory of multipole moments describing stress-energy tensors $T^{ab} = T^{(ab)}$ potentially coupled to an electromagnetic field $F_{ab}$. It was shown that -- quite surprisingly -- all degrees of freedom constrained by stress-energy conservation could be encoded in particular choices for the body's linear and angular momenta. The ordinary differential equations describing the evolution of these two quantities were shown to be equivalent to the partial differential equation
\begin{equation}
  \nabla^b T_{ab} = F_{ab} J^b ,
  \label{StressConsEM}
\end{equation}
where $J^a$ is the body's conserved electromagnetic current density. The time evolution of the quadrupole and higher moments is not constrained by stress-energy conservation\footnote{There are additional subtleties arising from the assumption that the moments describe a smooth stress-energy tensor. The form of the resulting constraints is not completely understood, although the analogous problem has been solved for the reduced multipole moments of $J^a$ \cite{CurrentMoments}.}. Knowing these moments together with the linear and angular momenta is sufficient to completely reconstruct a generic stress-energy tensor.

Both exact integrals and approximate multipole series were derived for the force and torque acting on an arbitrary charged object with finite radius. A center of mass worldline could also be defined from the momenta. The resulting laws of motion have had an important influence on the understanding of test body motion in general relativity. The formalism also applies to objects with strong self-fields, although it is then much more difficult to approximate the motion in a useful way.

Despite this, the consequences of Dixon's definitions have been successfully studied for a wide class of self-interacting extended charges in flat spacetime \cite{HarteEMOld}. The resulting electromagnetic self-force agreed with expectations only under very strong assumptions. It was later understood that the problematic terms were related to effective renormalizations of \textit{all} components of an object's linear and angular momenta \cite{HarteScalar}. This is much more difficult to recognize ``by inspection'' than the classical mass shifts appearing in simple cases like those described by \eqref{nonRelRR}.

Dixon's formalism has recently been modified in order to much more easily and naturally take into account these kinds of effects. All momentum renormalizations are included at the outset using arguments that do not involve any perturbative expansions of the self-field. Scalar and electromagnetic versions of this formalism have both been developed in fixed spacetimes \cite{HarteScalar, HarteEMNew}. Similar ideas have also been applied in general relativity linearized off of an arbitrary vacuum background \cite{HarteGrav}. Only the results of the scalar case will be reviewed here.

\subsection{Generalized symmetries}
\label{Sect: GKFs}

If exact Killing fields do not exist, it becomes convenient to define momenta associated with vector fields that are ``nearly Killing'' near a given worldline. Such objects require the specification of a frame; namely a timelike worldline $\Gamma$ together with a set of spacelike hypersurfaces $\{ \Sigma_s | \forall s \in \mathbb{R} \}$ foliating a worldtube $\tilde{W} \supset \Gamma$. Any point in a given $\Sigma_s$ is assumed to be connected to $\gamma_s = \Gamma \cap \Sigma_s$ by a unique geodesic lying entirely within that hypersurface.

Given such a frame, a set of generalized Killing fields (GKFs) may be defined within $\tilde{W}$. These share several properties with ordinary Killing fields. Most importantly, knowledge of a GKF and its (antisymmetric) first derivative at any point on $\Gamma$ fixes that vector field throughout $\tilde{W}$. The dependence on these initial data is linear and non-degenerate. For a given frame, there therefore exist exactly 10 linearly independent GKFs in every four dimensional spacetime. They satisfy
\begin{equation}
  \LieX g_{ab} |_\Gamma = \nabla_a \LieX g_{bc} |_\Gamma = 0.
  \label{AlmostKilling}
\end{equation}
Any ordinary Killing fields that may exist are also GKFs.

These were the only properties used in \cite{HarteScalar, HarteEMNew} in order to derive laws of motion for matter distributions coupled to their own scalar and electromagnetic fields. They do not provide a unique definition, however. In order to explicitly evaluate the self-force beyond its lowest order, the GKFs proposed in \cite{HarteSyms} will be adopted here\footnote{In \cite{HarteSyms}, a larger class of vector fields were introduced as generalized affine collineations (GACs). The objects needed here are special cases originally referred to as Killing-type GACs. Following \cite{HarteScalar, HarteEMNew}, we simplify the terminology and relabel these as GKFs.}. They are derived from a one-parameter family of Jacobi fields $\psi^a(x,s)$. Associated with each point $\gamma_s$ on $\Gamma$ are two tensors $\mathcal{A}_a(s)$ and $\mathcal{B}_{ab} = \mathcal{B}_{[ab]}(s)$ related to the Jacobi fields via
\begin{equation}
  \mathcal{A}_a(s) = \psi_a(\gamma_s,s) , \qquad  \mathcal{B}_{a b}(s) = \nabla_a \psi_b (\gamma_s,s).
  \label{InitData}
\end{equation}
Using these quantities as initial data, $\psi^a(x,s)$ may be found away from $\gamma_s$ by integrating
\begin{equation}
  \frac{\mathrm{D}^2 \psi_a}{\rmd u^2} - R_{abc}{}^{d} \dot{y}^b \dot{y}^c \psi_d = 0 \label{Jacobi}
\end{equation}
along an affinely-parameterized geodesic $y(u)$ connecting that point to $x$.

Eq. \eqref{Jacobi} is just the geodesic deviation or Jacobi equation. It is easily verified to be equivalent to the two first-order equations
\begin{equation}
  \frac{\mathrm{D} \psi_a }{ \rmd u } - \frac{\rmd y^b}{\rmd u} \hat{\mathcal{B}}_{ba} = 0, \qquad \frac{\rmd y^a}{\rmd u} \left( \frac{\mathrm{D} \hat{\mathcal{B}}_{ab}}{ \rmd u} + R_{abc}{}^{d} \frac{\rmd y^c}{\rmd u} \psi_d \right) = 0, \label{ProjectedKT}
\end{equation}
together with initial conditions $\psi_a(\gamma_s,s) = \mathcal{A}_a(s)$ and $\hat{\mathcal{B}}_{ab}(\gamma_s,s) = \mathcal{B}_{ab}(s)$. Both of these forms have simple interpretations. The first-order equations are very nearly the Killing transport equations that relate genuine Killing fields at different points. The only difference is the overall factor of $\dot{y}^a$ in the second half of \eqref{ProjectedKT}. Without this projection, the full Killing transport equations would lead to inconsistently propagated derivatives.

Next, let $\tau(x)$ be a ``time function'' that returns which leaf of the foliation its argument belongs to. This means that $\tau(x) = s$ for every $x \in \Sigma_s$. A GKF $\xi^a$ may now be derived from $\psi^a$ by setting
\begin{equation}
  \xi^a(x) = \psi^a(x,\tau(x)) . \label{XiDef}
\end{equation}
It may be shown that \eqref{AlmostKilling} implies
\begin{equation}
    \frac{\mathrm{D} \mathcal{A}_a }{ \rmd s } - \dot{\gamma}^b_s \mathcal{B}_{b a} = 0, \qquad \frac{\mathrm{D} \mathcal{B}_{ab}}{ \rmd s} + R_{a b c}{}^{d} \dot{\gamma}^c_s \mathcal{A}_d = 0 .\label{FullKT}
\end{equation}
These are the full Killing transport equations, and should be contrasted with \eqref{ProjectedKT}. They specify a unique family of Jacobi fields -- and therefore a unique GKF -- for any choice of $\mathcal{A}_a(s_0)$ and $\mathcal{B}_{ab} = \mathcal{B}_{[ab]}(s_0)$ given at some point $\gamma_{s_0} \in \Gamma$.

To summarize, a GKF may be obtained at any point in $\tilde{W}$ from initial data $\mathcal{A}_a(s_0)$ and $\mathcal{B}_{ab}(s_0)$. This is done by first integrating \eqref{FullKT} along an appropriate segment of $\Gamma$, and then \eqref{ProjectedKT} out to the point of interest. The resulting objects satisfy all of the properties discussed above. They are also related to the initial data via
\begin{equation}
  \xi_a (\gamma_s) = \mathcal{A}_a(s) , \qquad \nabla_a \xi_b (\gamma_s) = \mathcal{B}_{ab} (s) = \mathcal{B}_{[a b]}(s). \label{ABMeaning}
\end{equation}
Other consequences of this definition are discussed in \cite{HarteSyms}.

\subsection{Momentum}

Given some extended matter distribution with a spatially-compact worldtube $W \subseteq \tilde{W}$, we represent its total momentum with respect to the given frame by a linear functional $\ItP_\xi(s)$ of the GKFs at each point on $\Gamma$. If, for example, $Z^a = \partial / \partial Z$ were chosen to be a translational Killing field in some spacetime, $\ItP_Z$ would represent the $Z$-component of the body's linear momentum. It is more common to represent linear and angular momenta by tensor fields $p_a(s)$ and $S_{ab} = S_{[ab]}(s)$ on $\Gamma$. These objects are related to $\ItP_\xi$ via
\begin{equation}
  \ItP_\xi(s) = p_a(s) \xi^a (\gamma_s) + \frac{1}{2} S^{ab}(s) \nabla_{a} \xi_{b} (\gamma_s) .
  \label{TranslateMomenta}
\end{equation}
Working with the functional on the left-hand side provides a number of simplifications, although it is completely equivalent to using $p_a$ and $S_{ab}$.

Useful definitions for these objects have been derived in \cite{HarteScalar} and \cite{HarteEMNew} for matter coupled to scalar and electromagnetic fields, respectively. For simplicity, only massless, minimally-coupled scalar fields are considered here. The momenta then have the form\footnote{The notation here has changed somewhat from \cite{HarteScalar, HarteEMNew}. There, the first term of \eqref{PDef} by itself was referred to as $\ItP_\xi$. The sum was denoted by $\hat{\ItP}_\xi$.}
\begin{equation}
  \ItP_\xi = \int_{\Sigma_s} T^{a}{}_{b} \xi^b \rmd S_a + \mathcal{E}_\xi , \label{PDef}
\end{equation}
where $T_{ab}$ is the stress-energy of the body in question. $\mathcal{E}_\xi$ represents a contribution from the body's S-type self-field. It is defined by \eqref{EDef} below.

If the only source of energy or momentum immediately outside of the body is the scalar field $\phi$, the stress-energy tensor appearing in \eqref{PDef} is given by
\begin{equation}
  T_{ab} = T_{ab}^{\mathrm{tot}} - \frac{1}{4 \pi} ( \nabla_a \phi \nabla_b \phi - \frac{1}{2} g_{ab} \nabla_c \phi \nabla^c \phi)
\end{equation}
in a neighborhood of $W$. The total stress-energy tensor $T_{ab}^{\mathrm{tot}}$ is conserved:
\begin{equation}
  \nabla^a T_{ab}^{\mathrm{tot}} = 0.
  \label{StressCons}
\end{equation}

If $\rho$ is the body's scalar charge density, its self-field is defined to be the retarded solution to
\begin{equation}
  \nabla^a \nabla_a \phi = - 4 \pi \rho . \label{WaveScalar}
\end{equation}
The remaining field is labeled ``external;'' i.e.
\begin{equation}
  \phi = \phi_\rmExt + \phi_{\mathrm{ret}} = \phi_\rmExt + \phi_\rmSelf .
\end{equation}
This implies that $\phi_\rmExt$ is a solution to the homogeneous version of \eqref{WaveScalar} in a neighborhood of the body's worldtube. The scalar charge density $\rho$ is assumed to have similar support properties to $T_{ab}$. More precisely,
\begin{equation}
    \tilde{W} \supseteq W = \mathrm{supp} \,\, T_{ab} \supseteq \mathrm{supp} \, \, \rho .
\end{equation}

The self-field can be further split using a Detweiler-Whiting decomposition \cite{DetWhiting, PoissonRev}. Define S and R-fields satisfying
\begin{equation}
  \phi_{\mathrm{ret}} = \phiS + \phiR .
\end{equation}
The two quantities on the right-hand side are referred to as the singular (S) and regular (R) components of $\phi_\rmSelf$, respectively. Despite their names, both are mathematically well-behaved for physically reasonable charge distributions. Roughly speaking, $\phiS$ represents the ``Coulomb-like'' bound field and $\phiR$ the radiation. The latter object satisfies the homogeneous wave equation, and is usually responsible for most of the non-inertial self-force. These objects are defined via Green functions $\GR(x,x')$ and $\GS(x,x')$ using expressions like
\begin{equation}
  \phiR (x) = \int_W \rho(x') \GR (x,x') \rmd V' .
\end{equation}
They satisfy
\begin{equation}
  \nabla^a \nabla_a \GR = 0 , \qquad \nabla^a \nabla_a \GS = - 4\pi \delta(x,x'),
  \label{GreenWaveEqs}
\end{equation}
and
\begin{equation}
  \GS (x,x') = \GS (x',x).
\end{equation}
The S-type Green function also vanishes if its arguments are timelike-separated.

Two more brief definitions are necessary before writing down the self-field's contribution to a body's momentum. Bisect the body's worldtube with $\Sigma_s$. Let the halves to the future and past of this hypersurface be denoted by $\Sigma^{+}_s$ and $\Sigma^{-}_s$, respectively. Also define the field ``due to'' charge in a spacetime region $\Lambda$ by
\begin{equation}
  \phiS [\Lambda ](x) = \int_\Lambda \rho(x') \GS (x,x') \rmd V' .
\end{equation}
The scalar component of the self-momentum appearing in \eqref{PDef} is then \cite{HarteScalar}
\begin{equation}
  \mathcal{E}_\xi(s) = \frac{1}{2} \left( \int_{\Sigma^{+}_s} \rmd V \rho \LieX \phiS[\Sigma^{-}_s] - \int_{\Sigma^{-}_s} \rmd V \rho \LieX \phiS [\Sigma^{+}_s] \right).
  \label{EDef}
\end{equation}
Despite the unbounded volumes of $\Sigma^{+}_s$ and $\Sigma^{-}_s$, the support of $\GS$ restricts the integrals here to finite times into the past and future of order the body's diameter. These definitions reduce to standard results in stationary systems \cite{HarteScalar, HarteEMNew}. Regardless, the total momentum $\ItP_\xi$ in \eqref{PDef} has now been fixed in terms of $\rho$, $g_{ab}$, and $T_{ab}$.

\subsection{Laws of motion}

A body's bulk motion can be interpreted to mean the behavior of its momenta and an associated center of mass worldline. We therefore seek a method for determining $\ItP_\xi(s)$ at arbitrary times given appropriate initial data. The relevant differential equations follow from stress-energy conservation and the scalar wave equation \eqref{WaveScalar}.

Once they are known, rates of change for the ordinary momenta $p_a$ and $S_{ab}$ are easily extracted. Differentiating \eqref{TranslateMomenta} and applying \eqref{AlmostKilling} shows that
\begin{equation}
  \dot{\ItP}_\xi = ( \dot{p}_a + \frac{1}{2} R_{abcd} \dot{\gamma}^b_s S^{cd} ) \xi^a + \frac{1}{2} ( \dot{S}^{ab} - 2 p^{[a} \dot{\gamma}^{b]}_s ) \nabla_a \xi_b.
\end{equation}
It is clear from this that $\dot{\ItP}_\xi$ measures the degree to which the evolution of the body's momenta deviates from the Papapetrou equations describing the motion of a spinning (but otherwise structureless) test particle. For a fixed GKF $\xi^a$, it may be thought of as a particular linear combination of the net force and torque.

$\dot{\ItP}_\xi$ is derived in \cite{HarteScalar}. It will be useful here to split it into three parts. Let
\begin{equation}
  \dot\ItP_\xi= \ItF_\xi^{\mathrm{grav}} + \ItF_\xi^{\mathrm{hom}} + \ItF_\xi^\rmS .
  \label{PDot}
\end{equation}
The first term on the right-hand side represents the ``gravitational force'' (or torque). If $t^a$ is the time evolution vector field for the foliation $\{ \Sigma_s \}$, it is given by
\begin{equation}
  \ItF_\xi^{\mathrm{grav}} (s)  =  \frac{1}{2} \int_{\Sigma_s} T^{bc}(x) \LieX g_{bc}(x)  t^a (x) \rmd S_a.
  \label{FGrav}
\end{equation}
This couples to the quadrupole and higher moments of $T^{ab}$, and is a generalization of effects well-known even in Newtonian gravity.

Next, $\ItF_\xi^{\mathrm{hom}}$ denotes the ordinary force due to the scalar field $\phi_{\mathrm{hom}} = \phi_\rmExt + \phi_\rmR$:
\begin{equation}
  \ItF_\xi^{\mathrm{hom}}(s) = \int_{\Sigma_s} \rho(x) \LieX \phi_{\mathrm{hom}}(x) t^a(x) \rmd S_a .
  \label{FHom}
\end{equation}
Note that $\phi_{\mathrm{hom}}$ satisfies the homogeneous wave equation $\nabla^a \nabla_a \phi_{\mathrm{hom}} = 0$ in a neighborhood of $W$. In most cases, it therefore varies slowly inside the body. It is then possible to expand \eqref{FHom} in a multipole series about $\gamma_s$. To lowest order, this results in $\ItF_\xi^{\mathrm{hom}}(s) \simeq q \LieX \phi_{\mathrm{hom}}(\gamma_s)$. Standard results for the scalar self-force are then straightforward to obtain \cite{HarteScalar}.

Lastly, $\ItF_\xi^\rmS$ denotes the force exerted by the (inhomogeneous) S-type self-field. It is usually very small, as the vast majority of such effects have already been absorbed into the definition of $\ItP_\xi$. It was shown in \cite{HarteScalar} that
\begin{equation}
  \ItF_\xi^{\mathrm{S}}(s) = \frac{1}{2} \int_{\Sigma_s} \rmd S_a t^a(x) \int_W \rmd V' \rho(x) \rho(x') \LieX \GS(x,x') . \label{Def-SForce}
\end{equation}
Lie derivatives of two-point tensors are understood to act independently on each argument, so $\LieX \GS = \xi^a \nabla_a \GS + \xi^{a'} \nabla_{a'} \GS$. The remainder of this paper will focus on the consequences of \eqref{Def-SForce} and its similarity to the gravitational force $\ItF_\xi^{\mathrm{grav}}$.

This review has only discussed the evolution of a body's linear and angular momenta. The choice of frame $\{ \Gamma, \Sigma_s \}$ has been left open. One useful possibility involves introducing center of mass conditions \cite{Dix79,EhlRud}. The formalism presented here then leads directly to self-contained laws of motion in a more traditional sense. The details of such a process will not be needed below. The required steps are described in detail in various sources (see, e.g., \cite{HarteEMNew,Dix79,EhlRud}).

\section{Effective stress-energy tensors}
\label{Sect: EffMoments}

It follows from \eqref{Def-SForce} that the self-force and self-torque due to the S field are determined by $\LieX \GS$. The singular Detweiler-Whiting Green function is a purely geometric construction, so its Lie derivatives with respect to $\xi^a$ must be linear functionals of $\LieX g_{ab}$. This is also true of $\ItF_\xi^{\mathrm{S}}$. It may be seen more directly by using the field equation to replace $\rho(x)$ in \eqref{Def-SForce} by $-\nabla^a \nabla_a \phiS/4 \pi$, and then integrating by parts. This requires two intermediate identities. First, note that the second derivatives of any vector field can be written in the form \cite{HarteSyms}
\begin{equation}
  \nabla_b \nabla_a \xi_c = R_{cabd} \xi^d + \nabla_{(a} \LieX g_{b)c} - \frac{1}{2} \nabla_c \LieX g_{ab} .
  \label{DerivativeIdent}
\end{equation}
This generalizes a result that is  well-known for Killing fields (in which case the last two terms here vanish). It may be used to obtain a wave equation for $\LieX \GS$. Directly differentiating \eqref{GreenWaveEqs} shows that
\begin{eqnarray}
  \nabla^a \nabla_a \LieX \GS (x,x') &=& [ \nabla^a \nabla^b \GS + 2\pi \delta(x,x') g^{ab} ] \LieX  g_{ab} \nonumber
  \\
  && ~ + [ \nabla^b \LieX g_{ab} - \frac{1}{2} \nabla_a ( g^{bc} \LieX g_{bc} ) ] \nabla^a \GS .
  \label{BoxLie}
\end{eqnarray}
Combining this with \eqref{Def-SForce} gives
\begin{eqnarray}
  \ItF_\xi^\rmS (s) &=& \frac{1}{8 \pi} \int_{\Sigma_s} (\nabla^a \phiS \nabla^b \phiS - \frac{1}{2} g^{ab} \nabla^c \phiS \nabla_c \phiS ) \LieX g_{ab} t^c \rmd S_c \nonumber
  \\
  && \qquad ~ + \frac{\rmd}{\rmd s} \int_{\Sigma_s} \mathcal{I}^a_\xi \rmd S_a + \int_{\partial \Sigma_s} \mathcal{I}^a_\xi t^b \rmd S_{ab}  ,
  \label{SForceGen}
\end{eqnarray}
where
\begin{eqnarray}
  \mathcal{I}^a_\xi (x) &=& \frac{1}{8\pi} \int_W \rmd V' \rho' [ ( \nabla^a \phiS \LieX \GS - \phiS \nabla^a \LieX \GS) \nonumber
  \\
  && \qquad ~+ \phiS ( g^{ab} \nabla^c \GS - \frac{1}{2} g^{bc} \nabla^a \GS) \LieX g_{bc} ] .
\end{eqnarray}
The first line of \eqref{SForceGen} has a very simple interpretation. The quantity in parentheses is the stress-energy tensor of $\phiS$ (multiplied by $4 \pi$). Comparison with \eqref{FGrav} shows that there is a direct sense in which the stress-energy tensor of the body's self-field adds to that of its material components.

This is especially true in static, asymptotically flat systems. Temporarily suppose that there exists a timelike Killing field $T^a$, and that each $\Sigma_s$ is orthogonal to it. Also set $\Lie_T \rho = \Lie_T \phiS = 0$ and $\dot{\gamma}^a \propto T^a$. The time derivative in \eqref{SForceGen} clearly vanishes in this case. At large radii $r$, $\phiS$ decays like $1/r$. Similarly, there exist coordinate systems where the metric looks approximately Minkowski near spatial infinity: $g_{AB} \sim \eta_{AB} + \Or(1/r)$. Killing fields in flat spacetime can grow  like $r^1$. If the GKFs can be defined at large distances -- which is by no means guaranteed -- they might grow at similar rates. Assuming this to be true, $\LieX g_{AB} \sim \Or(r^0)$ and $\mathcal{I}^A_\xi \sim \Or(1/r^3)$. It follows that the integral over $\partial \Sigma_s$ in \eqref{SForceGen} will vanish if $\Sigma_s$ is pushed to spatial infinity. The only remaining term is the volume integral. It implies that the body responds to the background gravitational field as though it had a stress-energy tensor
\begin{equation}
  T^{ab}_{\mathrm{eff}} = T^{ab} + \frac{1}{4\pi} \left( \nabla^a \phiS \nabla^b \phiS - \frac{1}{2} g^{ab} \nabla^c \phiS \nabla_c \phiS \right) .
  \label{SelfStress}
\end{equation}
The extra terms here are exactly those expected for the stress-energy tensor of an (isolated) scalar field $\phiS$.

While suggestive, there are several problems with this result. First, it was only derived for static systems. It is not particularly clear that \eqref{SelfStress} can give the body's effective stress-energy tensor as it would be inferred by observing responses to different spacetime curvatures. As already alluded to, the derivation even in the static case is suspect due to both $\GS$ and the GKFs generically being undefined at very large distances. Further conceptual and technical problems also arise from the infinite extent of the stress-energy tensor associated with $\phiS$. It becomes very difficult to define effective multipole moments, for example. This is especially true if the spacetime is not Ricci-flat. Many of these problems can be at least partially circumvented at low orders in perturbation theory by using scaling arguments like those in \cite{BobSamMe}. It appears unlikely, however, that this kind of reasoning can be extended very far. No such attempt will be made here.

In general, either \eqref{Def-SForce} or \eqref{SForceGen} (with boundary terms) may be used to rigorously compute $\ItF_\xi^\rmS$. The latter equation was useful to motivate results like \eqref{SelfStress}, although it is simpler to use \eqref{Def-SForce} in calculations. The crucial point is to note -- as already mentioned -- that $\LieX \GS(x,x')$ is linearly dependent on $\LieX g_{ab}$ in some region near $x$ and $x'$. If the two arguments of this Green function lie sufficiently close together, it is well-known to take on the Hadamard form \cite{PoissonRev}
\begin{equation}
  \GS(x,x') = \frac{1}{2} \left[ \Delta^{1/2} \delta(\sigma) - V \Theta(\sigma) \right].
  \label{Hadamard}
\end{equation}
$\delta$ and $\Theta$ are the usual Dirac and Heaviside distributions, respectively. Synge's world function $\sigma(x,x') = \sigma(x',x)$ is defined in the appendix via \eqref{SigDef}. The scalarized van Vleck determinant is
\begin{equation}
  \Delta(x,x') = \Delta(x',x) = - \frac{\det ( - \sigma_{aa'} ) }{\sqrt{-g} \sqrt{-g'}},
  \label{Def-VanVleck}
\end{equation}
where we have used the standard notation $\sigma_{aa'} = \nabla_a \nabla_{a'} \sigma$. No similarly simple formula is known for the tail $V(x,x') = V(x',x)$. It is, however, a smooth biscalar determined entirely by the spacetime geometry near $x$ and $x'$.

Lie derivatives of $\sigma$, $\Delta$ and $V$ with respect to $\xi^a$ are all linear functionals of $\LieX g_{ab}$. It immediately follows from \eqref{GradSigma} that
\begin{equation}
  \LieX \sigma(x,x') = \frac{1}{2} \int_0^1 \rmd u \, \frac{\mathrm{D} y^a}{\rmd u} \frac{\mathrm{D} y^b}{\rmd u} \LieX g_{ab}(y) ,
  \label{LieSigmaBase}
\end{equation}
for example. As in \eqref{SigDef}, $y(u)$ is an affinely parameterized geodesic connecting $x$ and $x'$. Also note that a direct differentiation of \eqref{Def-VanVleck} gives
\begin{equation}
  \nabla_a \ln \Delta = - H^{b'}{}_{b} \sigma^{b}{}_{ab'} ,
  \label{GravDelta}
\end{equation}
where
\begin{equation}
  H^{a'}{}_{a}(x,x') = [-\sigma^{a}{}_{a'} (x,x') ]^{-1}.
  \label{Def-H}
\end{equation}
The ``$-1$'' in this last equation denotes a matrix inverse, which is always assumed to exist in the regions of interest. It immediately follows from \eqref{GravDelta} that
\begin{equation}
  \LieX \ln \Delta(x,x') = - [\nabla_a \xi^a + \nabla_{a'} \xi^{a'} + H^{a'a} \nabla_a \nabla_{a'} \LieX \sigma(x,x') ] .
  \label{LieDeltaBase}
\end{equation}
$2 \nabla_a \xi^a = g^{ab} \LieX g_{ab}$, so, with the help of \eqref{LieSigmaBase}, $\LieX \Delta$ can also be written entirely in terms of $\LieX g_{ab}$.

Such explicit demonstrations do not seem to be easily obtained for $V(x,x')$. Regardless, the tail is a smooth solution to the homogeneous wave equation: $\nabla^a \nabla_a V =0$. It can be uniquely specified via purely geometric boundary data given on a null cone with vertex $x$ or $x'$ \cite{PoissonRev}. It is therefore conceptually clear that $V$ is an entirely geometric quantity whose Lie derivatives may also be written in terms of $\LieX g_{ab}$. This is supported by direct Taylor expansions of $V$ that have been computed to fairly high order in the literature \cite{GreenExpansion1, GreenExpansion2}.

It is now clear from \eqref{Def-SForce} that $\ItF_\xi^\rmS(s)$ can be written as a linear functional of $\LieX g_{ab}$ as it appears in a certain finite neighborhood of $\Sigma_s \cap W$. Inspection of \eqref{FGrav} shows that the gravitational force $\ItF_\xi^{\mathrm{grav}}(s)$ is also linear in $\LieX g_{ab}$, although only \textit{on} $\Sigma_s \cap W$. Despite this difference, the two forces may still be related if the metric is analytic (in a Riemann normal coordinate system with origin $\gamma_s$). In this case, $\LieX g_{ab}$ can be expanded in a Taylor series about $\gamma_s$. As will be discussed in more detail in Sect. \ref{Sect: ExpandLie} below, every term will involve a quantity $\LieX g_{ab,c_1 \cdots c_n} (\gamma_s)$ for some $n \geq 2$. $g_{ab,c_1 \cdots c_n}$ is the $n^{\mathrm{th}}$ metric normal tensor (or tensor extension of $g_{ab}$), and is defined by \eqref{Def-MetricExt}.

Using the resulting expansion in \eqref{FGrav} or \eqref{Def-SForce} yields infinite series for $\ItF_\xi^{\mathrm{grav}}$ and $\ItF_\xi^\rmS$, respectively. A term involving $\LieX g_{ab,c_1 \cdots c_n} (\gamma_s)$ clearly must be contracted with a tensor of rank $n+2$. In the case of the gravitational force, this tensor is the $2^n$-pole moment $I^{c_1 \cdots c_n ab}(s)$ of $T^{ab}$ (divided by $2 \cdot n!$). See Sect. \ref{Sect: GravMoments} below. Similar coefficients in the series for $\ItF_\xi^\rmS$ may be identified as effective shifts in these multipole moments. This means that
\begin{equation}
  \ItF_\xi^{\mathrm{grav}}(s) + \ItF_\xi^\rmS (s) = \frac{1}{2} \sum_{n=2}^{\infty} \frac{1}{n!} I^{c_1 \cdots c_n ab}_{\mathrm{eff}} (s) \LieX g_{ab,c_1 \cdots c_n} (\gamma_s)
  \label{RenGravForce}
\end{equation}
for some effective multipole moments $I^{c_1 \cdots c_n ab}_{\mathrm{eff}} = I^{c_1 \cdots c_n ab} + \delta I^{c_1 \cdots c_n ab}$. This should be compared with the purely gravitational force \eqref{GravForce}. Formulae for the ``bare'' moments $I^{c_1 \cdots c_n ab}$ in terms of the matter fields may be found in the literature \cite{Dix74}, and are rederived here in Sect. \ref{Sect: GravMoments}. The shifts $\delta I^{c_1 \cdots c_n ab}$ are due to the self-field, and may be calculated from \eqref{Def-SForce}. Much of the remainder of this paper will be concerned with this process.

It should be noted that similar results are obtained if the metric is not analytic, but still varies slowly near $\Sigma_s \cap W$. In this case, it is often possible to write down asymptotic series for the gravitational force and torque that are very similar to those just described  \cite{Dix67}. The only difference is that the series becomes an approximation that must be cut off at finite order. The same comments also apply to $\ItF_\xi^\rmS$.

If the metric is not smooth near the body, there needn't be any relation between $\ItF_\xi^{\mathrm{grav}}$ and  $\ItF_\xi^\rmS$. Consider, for example, a spacetime that is everywhere flat except on a null hypersurface. As the gravitational wave approaches the body in question, there will be a very brief period during which the gravitational force exactly vanishes, yet $\ItF_\xi^\rmS \neq 0$. Momentum changes or center of mass motions predicted to arise on such short timescales are unlikely to have any physical relevance. This paper will therefore be concerned only with systems where $\LieX g_{ab}$ can be approximated by high order Taylor series in all regions of interest. The requisite smoothness assumptions are very similar to those used in the ordinary multipole expansion as it relates to test body motion. That problem has been discussed in detail by Dixon \cite{Dix67}.

\section{Expanding $\LieX g_{ab}$}
\label{Sect: ExpandLie}

A strong case has now been made that the singular scalar self-field mimics the ordinary effects of curvature felt even by extended test masses. We now explicitly expand $\LieX g_{ab}(x)$ as outlined above. This is fundamental to calculating the shifts in a body's multipole moments.

Many of the equations used in this section involve multiple spacetime points. Indicating which indices are associated with which points is notationally cumbersome. A more consistent use of primes will therefore be made in order to avoid confusion. Simpler equations will have primes removed when their meanings are clear. Points on $\Gamma$ will be denoted by symbols that are left unprimed in most cases.

If one is interested in finding a generalized force $\dot{\ItP}_\xi(s)$ at time $s$, it is useful to expand Lie derivatives of the metric about the point $\gamma_s \in \Gamma$. A series of this sort for $g_{ab}$ is given by \eqref{MetricExpand}. It involves ``separation vectors'' $X^a = - \sigma^a (x,x')$ and metric normal tensors $g_{ab,c_1 \cdots c_n}$. Lie derivatives may be computed directly. First note that
\begin{equation}
  \LieX g_{a'b'}(x') = \left[ \Lie_\psi g_{a'b'} + 2 \nabla_{(a'} \tau \frac{ \rmd }{ \rmd s} \psi_{b')} \right]_{ ( x',\tau(x') ) },
  \label{LieXvsLiePsi}
\end{equation}
which follows from \eqref{XiDef}. $\psi^{a'}(x',\tau(x'))$ is the Jacobi field from which $\xi^{a'}(x')$ is to be constructed. It is implicit that the second argument in this field is to be evaluated at $\tau(x')$ only after all differentiations have been performed. This means that
\begin{equation}
  \left[ \Lie_\psi g_{a'b'} \right]_{ ( x',\tau(x') ) } = 2 \left. \nabla_{(a'} \psi_{b')}(x',t) \right|_{t= \tau(x')},
\end{equation}
for example.

The first term in \eqref{LieXvsLiePsi} is easily evaluated using \eqref{MetricExpand} and the identity \cite{HarteSyms}
\begin{equation}
  \Lie_\psi \sigma^{a} (x', \gamma_s ) = \Lie_\psi \sigma^{a}{}_{b'} (x', \gamma_s) = 0 .  \label{LieSig}
\end{equation}
This notation is somewhat ambiguous regarding the arguments of the Jacobi fields. Fully expanding the left-hand side,
\begin{eqnarray}
  \Lie_\psi \sigma^{a} (x', \gamma_s ) &=& \psi^{b'}(x',s) \sigma^{a}{}_{b'}(x',\gamma_s) + \psi^b(\gamma_s,s) \sigma^{a}{}_{b}(x',\gamma_s) \nonumber
  \\
  && ~ - \nabla_b \psi^a(\gamma_s, s) \sigma^b(x',\gamma_s).
\end{eqnarray}
This is valid for any $s$. Combining these equations with \eqref{AlmostKilling} yields
\begin{equation}
  \Lie_\psi g_{a'b'}(x') \simeq \left[ \sigma^{a}{}_{a'} \sigma^{b}{}_{b'} \sum_{n=2}^N \frac{1}{n!} X^{c_1} \cdots X^{c_n} \LieX g_{a b,c_1 \cdots c_n}  \right]_{ (x',s )}.
  \label{LieGJacobi}
\end{equation}
From the perspective of \eqref{LieXvsLiePsi}, this is only needed in the special case $s = \tau(x')$. It is valid for general $s$, however.

Evaluating $\rmd \psi_{a'} / \rmd s$ in \eqref{LieXvsLiePsi} requires some additional care. A direct differentiation of \eqref{LieSig} with respect to $s$ while using \eqref{InitData} and \eqref{FullKT} shows that
\begin{equation}
  \frac{\rmd}{\rmd s} \psi^{a'} = \dot{\gamma}^{b}_s H^{a'}{}_{c} \Lie_\psi \sigma^{c}{}_{b} ,
  \label{PsiDotBase}
\end{equation}
where $H^{a'}{}_{a}$ is defined by \eqref{Def-H}. Also note that
\begin{equation}
  \sigma^{a}{}_{b} = \sigma^{c'} \sigma^{a}{}_{b c'} + \sigma^{a}{}_{c'} \sigma^{c'}{}_{b} ,
\end{equation}
which follows from \eqref{SigIdent}. If $y''(u)$  is an affinely parameterized geodesic connecting $\gamma$ to $x'$, both of these results together with \eqref{LieSig} imply the ``transport equation''
\begin{equation}
  0 = \left[ u^2 \frac{\rmd}{\rmd u} ( u^{-1} \Lie_\psi \sigma^{a}{}_{b}) - \sigma^{a c''} \sigma_{b}{}^{d''} \Lie_\psi g_{c''d''} \right]_{(y''(u),\gamma_s)} .
  \label{TransportLieSig2}
\end{equation}
Normalizing $u$ such that $y''(0) = \gamma_s$ and $y''(1) = x'$, an integral of this equation yields
\begin{eqnarray}
  \frac{\rmd}{\rmd s} \psi^{a'} (x',s) &=& \dot{\gamma}^{b}_s H^{a'c}(x',\gamma_s) \int_0^1 \rmd u u^{-2} \sigma_{c}{}^{ d''}(y'',\gamma_s)
  \nonumber
  \\
  && ~ \times \sigma_{b}{}^{f''}(y'',\gamma_s) \Lie_\psi g_{d''f''}(y'') .
  \label{PsiDot}
\end{eqnarray}
As with \eqref{LieGJacobi}, this applies for any $s$ (as long as $\gamma_s$ is not too far from $x'$). Making use of that equation here yields a series for $\rmd \psi^{a'}/\rmd s$ in terms of the metric normal tensors at $\gamma_s$.

Using the result together with \eqref{LieXvsLiePsi} and \eqref{LieGJacobi} finally gives
\begin{eqnarray}
  \fl \qquad \qquad \LieX g_{a'b'}(x') \simeq & \sum_{n=2}^N \frac{1}{n!} \bigg[ \left( \sigma^{a}{}_{a'} \sigma^{b}{}_{b'} + \frac{2}{n-1} \stackrel{(n)}{\Theta} \! {}^{a b}{}_{d f} \dot{\gamma}^{d}_s H_{(a'}{}^{f} \nabla_{b')} \tau \right) \nonumber
  \\
  & ~ \times X^{c_1} \cdots X^{c_n} \LieX g_{a b, c_1 \cdots c_n} \bigg]_{ ( x',\tau(x') ) },
  \label{LieGExpandFull}
\end{eqnarray}
where
\begin{equation}
  \fl \quad \qquad \stackrel{(n)}{\Theta} \! {}^{a b c d}(x',\gamma_s) = (n-1) \int_0^1 \rmd u u^{n-2} \left[ \sigma^{a f''} \sigma^{(c}{}_{f''} \sigma^{d)}{}_{h''} \sigma^{b h''} \right]_{ (y''(u),\gamma) }
\end{equation}
for all $n \geq 2$. In flat spacetime,
\begin{equation}
  \stackrel{(n)}{\Theta} \! {}^{a b c d} \rightarrow g^{a(c} g^{d) b} .
\end{equation}
These expressions succeed in expanding $\LieX g_{ab}(x)$ up to $N$th order about $\gamma_{\tau(x)}$. They generalize and unify results found (by direct computation) in \cite{HarteSyms}. They are all that is needed to expand $\ItF_\xi^{\mathrm{grav}}$ in a multipole series.

Unfortunately, computing $\ItF_\xi^\rmS$ requires knowledge of $\LieX g_{ab}(x)$ at points $x$ not contained in $\Sigma_{\tau(x)}$. Simple closed-form expressions do not seem easy to obtain in such cases. Despite this, it is still possible to find a useful approximation. Equation \eqref{LieGExpandFull} is of the form
\begin{equation}
  \LieX g_{a'b'}(x') \simeq \sum_{n=2}^N \frac{1}{n!} \stackrel{(n)}{\mathcal{G}}_{a'b'}(x',\tau(x')) ,
  \label{GExpParameter}
\end{equation}
where
\begin{eqnarray}
  \stackrel{(n)}{\mathcal{G}}_{a'b'}(x',s) &=& \bigg[ \left( \sigma^{a}{}_{a'} \sigma^{b}{}_{b'} + \frac{2}{n-1} \stackrel{(n)}{\Theta} \! {}^{a b}{}_{d f} \dot{\gamma}^{d}_s H_{(a'}{}^{f} \nabla_{b')} \tau \right) \nonumber
  \\
  && \times X^{c_1} \cdots X^{c_n} \LieX g_{a b, c_1 \cdots c_n} \bigg]_{ (x',s) }.
\end{eqnarray}
Note that the arguments here are $(x',s)$ rather than the $(x',\tau(x'))$ appearing in \eqref{LieGExpandFull}. Each of these coefficients may be further expanded as
\begin{equation}
  \stackrel{(n)}{\mathcal{G}}_{a'b'}(x',\tau(x')) \simeq \sum_{\ell=0}^L  \frac{(\tau(x')-s)^\ell }{\ell!} \stackrel{(n,\ell)}{\mathcal{G}}_{ab}(x',s)
  \label{GExp}
\end{equation}
for some $L \geq 0$. This will require $L$ derivatives of $\LieX g_{ab,c_1 \cdots c_n}(\gamma_s)$. Using \eqref{FullKT},
\begin{equation}
  \frac{\mathrm{D}}{\rmd s} \LieX g_{ab, c_1 \cdots c_n}(\gamma_s) = \dot{\gamma}^d_s \LieX \nabla_d g_{ab, c_1 \cdots c_n} (\gamma_s).
\end{equation}
The Thomas replacement theorem discussed in the appendix implies that $\nabla_d g_{ab, c_1 \cdots c_n}$ can be written in terms of (undifferentiated) metric normal tensors. It follows that $s$-derivatives of $\LieX g_{ab,c_1 \cdots c_n}$ are always proportional to Lie derivatives of other metric normal tensors. Carrying out all of the expansions here therefore leads to a series for $\LieX g_{a'b'}$ in which each term is proportional to some $\LieX g_{ab,c_1 \cdots c_n}(\gamma_s)$.

There does not appear to be any simple way to compute this exactly. It is, however, possible to obtain an approximation for $\stackrel{(n)}{\mathcal{G}}_{a'b'} (x',\tau(x'))$ that's valid to $n^{\mathrm{th}}$ order in powers of the separation between $x'$ and $\gamma_s$. Suppose that each hypersurface $\Sigma_s$ is formed by combining all geodesics passing through $\gamma_s$ and orthogonal to a unit timelike vector $n^a(s)$. If this satisfies $n_a \gamma_s^a = -1$, we have
\begin{equation}
  \nabla_a \tau(\gamma_s) = -n_a(s).
\end{equation}
Differentiating \eqref{GExp} and using standard results for the coincidence limits of various bitensors, it is now possible to show that
\begin{eqnarray}
  \fl \qquad \qquad \stackrel{(n)}{\mathcal{G}}_{a'b'}(x',\tau(x')) \simeq & \bigg[ \sigma^{(a}{}_{a'} \sigma^{c)}{}_{b'} \left( \delta^b{}_{c} - \frac{2}{n-1} \dot{\gamma}_s^b n_c \right) X^{d_1}_\bot \cdots X^{d_n}_\bot
  \nonumber
  \\
  & ~ \times  \LieX g_{ab, d_1 \cdots d_n} \bigg]_{( x',s )} + \Or(X^{n+1}) .
  \label{LieGApprox}
\end{eqnarray}
This only depends on the ``projected separation''
\begin{equation}
  X^a_\bot (x',\gamma_s) =  - [ \delta^{a}{}_{b} + \dot{\gamma}^a_s n_b (s) ] \sigma^b (x',\gamma_s)
\end{equation}
from the origin $\gamma_s$. In conjunction with \eqref{GExpParameter}, it will be fundamental to estimating the effective shift in the quadrupole ($n=2$) moment of $T^{ab}$.

\section{Multipole expansions}
\label{Sect: Quadrupole}

\subsection{Gravitational force and torque}
\label{Sect: GravMoments}

As claimed above, it follows from \eqref{FGrav} and \eqref{LieGExpandFull} that the gravitational force and torque may be expanded in a series of the form
\begin{equation}
  \ItF_\xi^{\mathrm{grav}}(s) \simeq \frac{1}{2} \sum_{n=2}^N \frac{1}{n!} I^{c_1 \cdots c_n a b}(s) \LieX g_{a b,c_1 \cdots c_n}(\gamma_s) .
  \label{GravForce}
\end{equation}
$I^{c_1 \cdots c_n a b}(s)$ represents the $2^n$-pole moment of the body's stress-energy tensor at time $s$. Given \eqref{MetricExtSyms0} and \eqref{MetricExtSyms}, it must satisfy
\begin{equation}
  I^{c_1 \cdots c_n ab} = I^{(c_1 \cdots c_n) ab} = I^{c_1 \cdots c_n (ab)},
\end{equation}
and
\begin{equation}
  I^{(c_1 \cdots c_n a) b} = I^{c_1 (c_2 \cdots c_n ab)} = 0.
\end{equation}

Following \cite{Dix74,Dix67}, it is convenient to define an alternative set of moments
\begin{equation}
  J^{f_1 \cdots f_n abcd} = I^{f_1 \cdots f_n [a[cb]d]}.
  \label{JDef}
\end{equation}
The notation on the right-hand side indicates independent (binary) antisymmetrizations with respect to the index pairs $(a,b)$ and $(c,d)$. It immediately follows from this definition that
\begin{equation}
  J^{f_1 \cdots f_n abcd} = J^{f_1 \cdots f_n [ab]cd} = J^{f_1 \cdots f_n ab[cd]} = J^{(f_1 \cdots f_n) abcd}.
  \label{JSyms1}
\end{equation}
Using the properties of $I^{c_1 \cdots c_n ab}$, we also have
\begin{equation}
  J^{f_1 \cdots f_n a [bcd]} = J^{f_1 \cdots f_{n-1} [f_n ab] cd} = 0.
  \label{JSyms2}
\end{equation}
This allows \eqref{JDef} to be inverted. Explicitly, \cite{Dix74,Dix67}
\begin{equation}
  I^{f_1 \cdots f_n ab} = 4 \left( \frac{n-1}{n+1} \right) J^{(f_1 \cdots f_{n-1} |a| f_n) b}
  \label{JToI}
\end{equation}
for all $n \geq 2$.

Using these relations together with \eqref{LieGExpandFull} and \eqref{GravForce} gives an explicit formula relating the multipole moments to $T^{ab}$:
\begin{eqnarray}
  \fl \qquad \qquad J^{f_1 \cdots f_n abcd}(s) = \int_{\Sigma_s} \rmd S_{h'} t^{h'} T^{a'b'} X^{f_1} \cdots X^{f_n} X^{[a} X^{[c}
  \nonumber
  \\
  ~ \times \left( \sigma^{b]}{}_{a'} \sigma^{d]}{}_{b'} + \frac{2}{n+1} \stackrel{(n+2)}{\Theta} \! {}^{b]d]}{}_{hl} \dot{\gamma}^{h}_s H_{(a'}{}^{l} \nabla_{b')} \tau \right).
  \label{StressEnergyMultipoles}
\end{eqnarray}
This result has also been derived in \cite{Dix74} using different methods. It is easily verified that \eqref{JSyms1} and \eqref{JSyms2} are satisfied. There is an additional identity
\begin{equation}
  n_{f_1} J^{f_1 \cdots f_n abcd} = n_{f_1} I^{f_1 \cdots f_n [a[cb]d]} = 0
\end{equation}
that also holds for all $n \geq 1$.

\subsection{Self-force due to the singular field}

Applying the same analysis to $\ItF_\xi^\rmS$ is more complicated. It requires $\stackrel{(n)}{\mathcal{G}}_{ab}(x,\tau(x))$, which is not known exactly in terms of metric normal tensors at $\gamma_s$. An approximation in powers of distance (divided by curvature scale) was necessary in order to obtain \eqref{LieGApprox}. We shall therefore adopt a similar small-size assumption for the matter under consideration. It is also useful to suppose that all timescales associated with the body's evolution are much larger than its diameter.

The singular self-force arises from the degree to which $\GS$ fails to remain invariant under the action of various GKFs. $\LieX \GS$ may be computed to lowest nontrivial order by making use of its Hadamard form \eqref{Hadamard}. Combining \eqref{GExpParameter}, \eqref{LieGApprox}, and \eqref{MetricExtLinear}, $\LieX g_{ab}$ is seen to reduce to
\begin{equation}
  \fl \qquad \quad \LieX g_{a'b'}(x') = \frac{1}{3} \sigma^{(a}{}_{a'} \sigma^{c)}{}_{b'} ( \delta^{b}{}_{c} - 2 \dot{\gamma}^b_s n_c ) X^d_\bot X^f_\bot \LieX R_{adfb}(\gamma_s) + \Or(X^3) .
\end{equation}
Substituting this into \eqref{LieSigmaBase} gives
\begin{eqnarray}
  \fl \LieX \sigma(x,y) = - \frac{1}{6} \{ X_\bot^a Y_\bot^b X_\bot^c Y_\bot^d + 2 \dot{\gamma}^a_s (X+Y)^b X^c_\bot Y^d_\bot [n_f (Y-X)^f] + \dot{\gamma}^a \dot{\gamma}^c_s \nonumber
  \\
  ~  \times (X^b X^d + X^{(b} Y^{d)} + Y^b Y^d ) [n_f (Y-X)^f]^2 \}\LieX R_{abcd} + \Or(X^5) ,
  \label{LieSigExpand}
\end{eqnarray}
where $X^a = - \sigma^a(x,\gamma_s)$ and $Y^a = - \sigma^a(y,\gamma_s)$. All indices here are associated with $\gamma_s$.

A similar calculation using \eqref{LieDeltaBase} gives
\begin{equation}
  \LieX \ln \Delta(x,y) = \frac{1}{6} (Y-X)^a (Y-X)^b \LieX R_{a b} + \Or(X^3).
  \label{LieDeltaExpand}
\end{equation}
$\LieX V$ is only needed to zeroth order in $X^a$. It may be found using the well-known coincidence limit \cite{PoissonRev}
\begin{equation}
  \lim_{y \rightarrow x} V(x,y) = \frac{1}{12} R(x) .
\end{equation}
An application of Synge's rule (see, e.g., \cite{PoissonRev}) to this equation shows that
\begin{equation}
  \LieX V(x,y) = \frac{1}{12} \LieX R(\gamma_s) + \Or(X) .
  \label{LieV}
\end{equation}
Together, \eqref{Hadamard}, \eqref{LieSigExpand}, \eqref{LieDeltaExpand}, and \eqref{LieV} specify $\LieX \GS$ to lowest nontrivial order.

Substituting the result into \eqref{Def-SForce} and performing the relevant integrations becomes significantly simpler if a more precise approximation is adopted for the time evolution of the matter distribution. Fix the orthonormal tetrad $e^a_A$ introduced in the appendix such that $e^a_0(\gamma_s) = n^a(s)$. Also suppose that the remaining spatial triad $e^a_I$ is Fermi-Walker transported along $\Gamma$. For all $I=1,2,3$, let
\begin{equation}
  \left( \frac{\mathrm{D}}{\rmd s} e^a_{I} + 2 \dot{n}^{[a} n^{b]} g_{bc} e^{c}_{I} \right)_\Gamma = 0 .
\end{equation}
This allows the introduction of a scaled ``radial separation''
\begin{equation}
  \ItX^I(\lambda,x) =  - e_a^I(\tau(x)) \sigma^a ( x, \gamma_{\tau(x)})/\lambda.
\end{equation}
The parameter $\lambda>0$ is introduced for later convenience. A time coordinate for a point $x$ may be identified as $\tau(x)$. The four numbers $\tau$ and $\lambda \mathcal{X}^I$ then define ``pseudo-Fermi'' normal coordinates with respect to the frame $\{ \Gamma, \Sigma_s \}$\footnote{An ordinary Fermi normal coordinate system uses a foliation orthogonal to its central worldline. A more general choice is allowed here. The extra complication is necessary in order to properly define a center of mass frame.}.

Now consider a 1-parameter family of matter distributions $\{\rho , T^{ab} \}$ described by the parameter $\lambda$. Let these satisfy the scaling relations
\begin{eqnarray}
  \rho (\lambda, x) = \lambda^{-2 \alpha} \tilde{\rho}(\lambda, \tau, \ItX^I), \label{ScaleRho}
  \\
  T^{AB} (\lambda, x) = \lambda^{-2 \beta} \tilde{T}^{AB}(\lambda, \tau, \ItX^I ) , \label{ScaleT}
\end{eqnarray}
in the Fermi normal coordinate system just described. $\tilde{\rho}$ and $\tilde{T}^{AB}$ are assumed to be smooth in all of their arguments, and to have compact support in $\mathcal{X}^I$. $\alpha$ and $\beta$ are real constants. The given relations state that the stress-energy tensor and charge density both scale down to $\Gamma$ in a self-similar way as $\lambda \rightarrow 0$. It should be stressed that this is an assumption on the system's behavior in \textit{spacetime}. While it is almost always possible to choose initial data with some type of spatial self-similarity, time evolution needn't preserve it. Most importantly, \eqref{ScaleRho}-\eqref{ScaleT} imply that the body's internal timescales do not scale with their spatial extent. It should be kept in mind that this is demonstrably false in many interesting systems. The problem is made much worse for certain values of $\alpha$ and $\beta$. A sharp existence theorem would be difficult to obtain, however.

We shall simply assume that reasonable choices have been made. Note, however, that the body's proper diameter $D$ scales like $\lambda$ as $\lambda \rightarrow 0$. Similarly, its scalar charge will be proportional to $\lambda^{3-2 \alpha}$. The (bare) mass scales like $\lambda^{3-2\beta}$. These estimates may be used to reduce the parameter space in several ways. First assume that the potentials remain everywhere finite as $\lambda \rightarrow 0$. This implies that
\begin{equation}
  \alpha \leq 1 .
\end{equation}
Also suppose that the center of mass acceleration does not diverge in the small-$\lambda$ limit we are considering. This means that the charge-to-mass ratio should remain bounded. Hence,
\begin{equation}
  \alpha \leq \beta .
\end{equation}
Both of these inequalities together imply that the body's inertia is not predominately due to its self-fields. The case $\alpha=\beta =1$ was considered in \cite{HarteEMNew,BobSamMe}. Except for assuming the given inequalities, no particular choice will be made here.

The singular self-force may now be computed to lowest nontrivial order. Substituting the above matter assumptions and the expansion for $\LieX \GS$ into \eqref{Def-SForce} gives
\begin{eqnarray}
  \fl \ItF_\xi^\rmS(s) \simeq \frac{\lambda^{7-4\alpha}}{12} \int \rmd^3 \ItX \rmd^3 \ItY \frac{ \tilde{\rho}(\lambda, s, \ItX) \tilde{\rho}(\lambda, s,\ItY) }{\|\ItX - \ItY\|^3 } \bigg\{ n^a n^c \| \ItX -\ItY \|^2 \Big[ g^{bd} \| \ItX -\ItY \|^2
  \nonumber
  \\
  ~  - e^b_I e^d_J ( \ItX^I \ItX^J + \ItX^I \ItY^J + \ItY^I \ItY^J ) \Big] + e^a_I e^c_K \Big[e^b_J e^d_L \ItX^I \ItY^J \ItX^K \ItY^L
  \nonumber
  \\
  ~ + \frac{1}{2} g^{bd} \| \ItX - \ItY \|^2 \Big( (\ItX - \ItY)^I (\ItX - \ItY)^K - \delta^{IK} \| \ItX - \ItY \|^2 \Big) \Big] \bigg\}
  \nonumber
  \\
  ~ \times \LieX R_{abcd}(\gamma_s) + \Or(\lambda^{8-4 \alpha}) .
  \label{SForceFinal}
\end{eqnarray}
The notation $\| \ItX \|^2$ denotes the usual Euclidean norm $\delta_{IJ} \ItX^I \ItX^J$. To quadrupole order, the gravitational force \eqref{GravForce} reduces to
\begin{equation}
  \ItF_\xi^{\mathrm{grav}} \simeq -\frac{1}{6} J^{abcd} \LieX R_{abcd} .
  \label{QuadForce}
\end{equation}
$J^{abcd} \sim O(\lambda^{5-2\beta})$ is defined via \eqref{StressEnergyMultipoles} above. Comparison of \eqref{SForceFinal} and \eqref{QuadForce} shows that an object moves as though it had a stress-energy quadrupole moment
\begin{equation}
  J^{abcd}_{\mathrm{eff}} = J^{abcd} + \delta J^{abcd},
  \label{JEff}
\end{equation}
where
\begin{eqnarray}
  \fl \delta J^{abcd} =  \frac{\lambda^{7-4\alpha}}{2} \int \rmd^3 \ItX \rmd^3 \ItY \frac{ \tilde{\rho}(\lambda, s, \ItX) \tilde{\rho}(\lambda, s,\ItY) }{\|\ItX - \ItY\|^3 } \bigg\{ n^{[a} n^{[c} \| \ItX -\ItY \|^2 \Big[ e^{b]}_I e^{d]}_J ( \ItX^I \ItX^J
  \nonumber
  \\
  ~  + \ItX^I \ItY^J + \ItY^I \ItY^J ) - g^{b]d]} \| \ItX -\ItY \|^2 \Big] - e^{[a}_I e^{[c}_K \Big[e^{b]}_J e^{d]}_L \ItX^I \ItY^J \ItX^K \ItY^L
  \nonumber
  \\
  ~ + \frac{1}{2} g^{b]d]} \| \ItX - \ItY \|^2 \Big( (\ItX - \ItY)^I (\ItX - \ItY)^K - \delta^{IK} \| \ItX - \ItY \|^2 \Big) \Big] \bigg\}
  \nonumber
  \\
  ~ + \Or(\lambda^{8-4\alpha}).
  \label{DeltaJ}
\end{eqnarray}
It is easily verified that this satisfies the same index symmetries as the ``bare'' moment $J^{abcd}$:
\begin{eqnarray}
  \delta J^{abcd} = \delta J^{[ab]cd} = \delta J^{ab[cd]}
  \\
  \delta J^{a[bcd]} = \delta J^{[abc]d} = 0.
\end{eqnarray}

The physical content of \eqref{DeltaJ} is far from clear as it stands. It is useful to split the full quadrupole moment into three pieces $\delta Q^{ab} = \delta Q^{(ab)}$, $\delta \pi^{abc} = \delta \pi^{a[bc]}$ and $\delta S^{abcd} = \delta S^{[ab]cd} = \delta S^{ab[cd]}$. Let \cite{EhlRud}
\begin{equation}
  \delta J^{abcd} = - 3 n^{[a} \delta Q^{b][c} n^{d]} - n^{[a} \delta \pi^{b]cd} - n^{[c} \delta \pi^{d]ab} + \delta S^{abcd}.
\end{equation}
This decomposition is unique if each tensor here is assumed to be fully orthogonal to $n^a$. $\delta Q^{ab}$, $\delta \pi^{abc}$, and $\delta S^{abcd}$ are then interpreted as shifts in the body's mass, momentum, and stress quadrupoles, respectively. In general,
\begin{equation}
  \delta \pi^{[abc]} = \delta S^{[abc]d} = \delta S^{a[bcd]} = 0.
\end{equation}

Considerable simplifications occur if $\LieX R_{ab}(\gamma_s) =0$. This will almost always be the case in general relativity if -- as we have been assuming -- the body's own metric perturbations are ignored. The introduction of a cosmological constant does not change this situation since $\LieX g_{ab}(\gamma_s)=0$ by \eqref{AlmostKilling}. Assuming that the Ricci tensor can be ignored, traces of the three quadrupole moments decouple from the laws of motion. It is therefore useful to define trace-free versions of these tensors satisfying
\begin{equation}
  g_{ab} \delta Q^{ab}_{\mathrm{TF}}  = g_{ab} \delta \pi^{abc}_{\mathrm{TF}} = g_{ac} \delta S^{abcd}_{\mathrm{TF}} = 0.
\end{equation}
The last equation forces $\delta S^{abcd}_{\mathrm{TF}}$ to have all of the same algebraic symmetries as a three-dimensional Weyl tensor. It therefore vanishes. A ``trace-free'' version of the full quadrupole moment may be defined as
\begin{equation}
    \delta J^{abcd}_{\mathrm{TF}} = - 3 n^{[a} \delta Q^{b][c}_{\mathrm{TF}} n^{d]} - n^{[a} \delta \pi^{b]cd}_{\mathrm{TF}} - n^{[c} \delta \pi^{d]ab}_{\mathrm{TF}} .
\end{equation}
Despite the name, $g_{ac} \delta J^{abcd}_{\mathrm{TF}} \neq 0$. It is, however, much simpler than $\delta J^{abcd}$. Forces and torques computed using both moments are the same. Letting $C_{abcd}$ denote the Weyl tensor,
\begin{equation}
  \delta J^{abcd} \LieX R_{abcd} = \delta J^{abcd}_{\mathrm{TF}} \LieX C_{abcd}
\end{equation}
when $\LieX R_{ab} = 0$.

Equation \eqref{DeltaJ} can be used to show that $\delta \pi^{abc}_{\mathrm{TF}} = \Or (\lambda^{8-4\alpha})$ and
\begin{equation}
  \fl \qquad \qquad \delta Q^{ab}_{\mathrm{TF}} = \frac{\lambda^{7-4\alpha}}{2} \int \rmd^3 \ItX \tilde{\rho} \phiS e^a_I e^b_J (\ItX^I \ItX^J - \frac{1}{3} \delta^{IJ} \| \ItX \|^2 ) + \Or(\lambda^{8-4\alpha}) .
  \label{MassMoment}
\end{equation}
This is exactly the quadrupole moment that might have been guessed from elementary considerations together with an assumption that the effective stress-energy tensor was given by \eqref{SelfStress}. The same argument could not have been used to deduce the full quadrupole moment \eqref{DeltaJ} needed in non-vacuum spacetimes. In that case, attempting to compute full (finite-trace) moments of the field's stress-energy tensor would lead to divergent integrals. The approach taken here cannot be bypassed in general. This might be an important point when analyzing motion in alternative theories of gravity.

We have focused on deriving the correction to an object's quadrupole moment. The results of Sect. \ref{Sect: ExpandLie} can be used to derive the lowest-order shifts in the octupole and higher multipole moments as well. The main obstacle is a lack of knowledge regarding the behavior of the tail $V(x,x')$. Results are available in the literature up to fairly high (but finite) order \cite{GreenExpansion1, GreenExpansion2}. These could be used to compute shifts in the next few multipole moments above the quadrupole. The situation is considerably simpler if $R_{ab} =0$. In this case, an expansion for the tail will consist of powers of $X^a$ contracted against a polynomial in the Weyl tensor and its derivatives. It is impossible to construct any such terms that are linear in the curvature, so $V \sim \Or (C^2)$ in this case. Such terms would all be higher order than those due to $\LieX \sigma$ and $\LieX \Delta$ in the approximations adopted here. In vacuum, the tail can therefore be ignored when computing the multipole shifts to lowest nontrivial order.

\section{Discussion}

We have analyzed the contribution of the singular field to the laws of motion describing an extended scalar charge in curved spacetime. In most cases, the total force and torque is given by
\begin{equation}
  \dot{\ItP}_\xi \simeq \int_{\Sigma_s} \rho \LieX (\phi_{\mathrm{ext}} + \phiR)t^a \rmd S_a +  \frac{1}{2} \sum_{n=2}^{N} \frac{1}{n!} I^{c_1 \cdots c_n ab}_{\mathrm{eff}} \LieX g_{ab,c_1 \cdots c_n}  .
  \label{FinalForce}
\end{equation}
The multipole series here is asymptotic in general. The maximum value one may choose for $N$ depends only on the degree to which the metric may be accurately approximated by a Taylor series inside the body. If it is analytic near $\Sigma_s \cap W$ in a Riemann normal coordinate system with origin $\gamma_s$, $N$ may be taken as infinite. The right-hand side of \eqref{FinalForce} then becomes exact (with the usual caveat that any effects of the body on the metric have been ignored). As noted following \eqref{FHom}, the integral involving external and R-type scalar fields can usually be expanded in its own multipole series. The monopole term is all that is necessary to recover standard results for the self-force on a scalar charge \cite{HarteScalar}.

Note that all traces of the S field have been absorbed into the effective (or renormalized) multipole moments $I^{c_1 \cdots c_n ab}_{\mathrm{eff}}$. Methods to compute these have been discussed in detail in Sects. \ref{Sect: ExpandLie} and \ref{Sect: Quadrupole}. An approximation for the quadrupole moment is explicitly given by \eqref{JEff} with \eqref{JToI}, \eqref{StressEnergyMultipoles}, and \eqref{DeltaJ}. The resulting expression is simplified significantly if $\LieX R_{ab} = 0$ on the body's worldline. In this case, the force due to the S field is approximately
\begin{equation}
  \ItF_\xi^\rmS \simeq \frac{1}{2} n^a \delta Q^{bc}_{\mathrm{TF}} n^d \LieX C_{abcd},
\end{equation}
where $\delta Q^{bc}_{\mathrm{TF}}$ is a symmetric, trace-free tensor given by \eqref{MassMoment}. This is exactly what one would expect by applying a Newtonian quadrupole definition to the energy density of a static S field. No similarly simple interpretation exists when $\LieX R_{ab} \neq 0$, however.

As discussed in the introduction, it is common to treat the motion of a small body as though all observations of its behavior are ``external.'' Masses are then inferred by observing the change in acceleration in response to a known change in an external force, for example. The same is also true for the spin angular momentum and all higher moments of the body's stress-energy tensor. The laws of motion satisfied by an extended test particle contain a number of parameters such as these. They may all be fit to the observed variations in the motion of a real object as it experiences various perturbing influences. The central result of this paper states that once all such fitting is complete, there is only one effect that has not been taken into account for a body with large self-fields. That is the force and torque exerted by $\phiR$. In a sense, the (``interesting'') self-force and self-torque is given entirely by the regular component of the self-field. This is much smaller than the full self-field. The fact that it satisfies a vacuum field equation also means that it usually varies much more slowly than $\phi_\rmSelf$. This situation is very similar to the one usually assumed from the start in effective field theory treatments of the self-force problem (see, e.g. \cite{EffectiveFields1, EffectiveFields2}).

It also serves as a direct derivation of a generalized Detweiler-Whiting axiom. Our results have proven a slightly modified version of this conjecture to be accurate far beyond any regime in which it was originally expected to hold. Up to effects that renormalize test particle parameters, there exist linear and angular momenta definable for essentially arbitrary extended charges whose instantaneous evolution depends on the self-field only through $\phiR$. The correct momenta are exactly those introduced in \cite{HarteScalar} and reviewed in Sect. \ref{Sect: Review} above. Recall that the original proposal was intended to treat only small ``pointlike'' objects to lowest order in some approximation scheme \cite{DetWhiting}. In that context, the S field was assumed only to renormalize the body's inertial mass. Despite the vast improvement in generality accomplished here, the precise prescription for splitting up the self-field has not changed.

While this paper has focused on the case of a scalar charge moving in a fixed curved spacetime, its basic conclusions are easily generalized. The only essential point was that the singular self-force depended on $\LieX \GS$, where $\GS$ was a Green function defined only using geometric objects. In electromagnetism, the analog of \eqref{Def-SForce} is \cite{HarteEMNew}
\begin{equation}
  \ItF_\xi^\rmS = \frac{1}{2} \int_{\Sigma_s} \rmd S_a t^a \int_W \rmd V' J_b J_{b'} \LieX G_{\mathrm{S}}^{bb'} ,
\end{equation}
where $J_a$ is the charge's current density. $G_{\mathrm{S}}^{aa'}$ is the singular Detweiler-Whiting Green function, which is again defined purely in terms of the geometry. The situation is slightly more subtle when studying gravitational self-forces, although the resulting expression for the singular self-force is still very similar\footnote{The methods used here depend on the linearity of the field equations. The formalism developed in \cite{HarteGrav} therefore treats gravitational self-fields as perturbations satisfying the linearized Einstein equation off of some given vacuum background. More general results might be possible by iteratively linearizing, or assuming approximate stationarity or axisymmetry. This is not yet known. There is also, at present, no formalism that allows both gravitational and scalar (or electromagnetic) self-fields to be large simultaneously.}  \cite{HarteGrav}. The arguments presented in this paper here are easily extended to all of these cases.

The calculations carried out in Section \ref{Sect: Quadrupole} may be of observational interest if extended to the gravitational case. In that context, future interferometers might be able to precisely measure the effects of a neutron star's quadrupole moment on the gravitational waves emitted as it spirals into a supermassive black hole. They may also be able to do the same for the star's spin. Such measurements could allow one to learn something about the star's equation of state. This would require a precise relation between the quadrupole moment extracted from the gravitational wave signal and the star's stress-energy tensor. It would be interesting to compare the post-Newtonian expectations for this to those derived from the present formalism. The nature of the ``effacement principle'' derived here might also be compared to the ones discussed in post-Newtonian contexts \cite{DSX1, DSX2, DSX3, Kopeikin}.

\appendix

\setcounter{section}{0}

\section{Riemann normal coordinates and metric normal tensors}
\label{App: RNC}

A spacetime's world function (or Synge's function) $\sigma(x,x') = \sigma(x',x)$ is defined to be one-half of the squared geodesic distance between its arguments. Let $y(u)$ be the unique affinely-parameterized geodesic connecting two nearby points $x$ and $x'$. If $y(0)=x$ and $y(1)=x'$, the world function is given by
\begin{equation}
  \sigma(x,x') = \frac{1}{2} \int_0^1  \rmd u \, \frac{\mathrm{D} y^a}{\rmd u} \frac{\mathrm{D} y^b}{\rmd u} g_{ab}(y) . \label{SigDef}
\end{equation}
Multiple derivatives of these objects occur frequently, so it is conventional to express them by appending indices: $\nabla_a \nabla_{b'} \nabla_c \sigma(x,x')$ is shortened to $\sigma_{cb'a}(x,x')$, for example.

First derivatives of $\sigma$ are particularly interesting. They lie tangent to $y(u)$. Specifically,
\begin{equation}
  \nabla_a \sigma(x,x') = - \left. \frac{\mathrm{D} y_a}{\rmd u} \right|_{u=0} , \, \qquad \nabla_{a'} \sigma(x,x') =  \left. \frac{ \mathrm{D} y_{a'}}{\rmd u} \right|_{u=1}. \label{GradSigma}
\end{equation}
These objects therefore serve as useful ``separation vectors.'' Let $e^{a}_A(x)$ (with $A=0,\ldots,3$) be an orthonormal tetrad at a particular point $x$. In a Riemann normal coordinate system with origin $x$, an arbitrary nearby point $x'$ is associated with the coordinate values
\begin{equation}
  X^A(x,x') = - e^A_{a}(x) \sigma^{a}(x,x') .
  \label{Def-RNC}
\end{equation}
Differentiating this with respect to $x'$ shows that
\begin{equation}
  \nabla_{a'} X^A = - e_{a}^A \sigma^{a}{}_{a'} .
\end{equation}
The left-hand side here clearly reduces to the identity matrix when evaluated in Riemann normal coordinates with origin $x$. It follows that $-e_{a}^A \sigma^{a}{}_{a'}$ is a covariant way of writing the identity. $H^{a'}{}_{a} e^{a}_{A}$ also corresponds to the identity in Riemann normal coordinates if $H^{a'}{}_{a}$ is defined by \eqref{Def-H}.

The metric in these coordinates can therefore be written as a two-point tensor field
\begin{equation}
  G_{AB} (x,x') = e_A^{a}(x) e_B^{b}(x) H^{a'}{}_{a}(x,x') H^{b'}{}_{b}(x,x') g_{a'b'}(x') .
  \label{Def-RNCMetric}
\end{equation}
There is no obstacle to rewriting this object with arguments $(x,X^A)$ rather than $(x,x')$. Set
\begin{equation}
  \tilde{G}_{AB}(x, X^C(x,x')) = G_{AB}(x,x').
\end{equation}
The $N$th-order ``covariant Taylor series'' of $G_{AB}$ is then
\begin{equation}
  G_{AB}(x,x') = \sum_{n=0}^{N} \frac{1}{n!}  X^{C_1} \cdots X^{C_n} \left[ \frac{\partial^n \tilde{G}_{AB}}{\partial X^{C_1} \cdots \partial X^{C_n}} \right]_{(x,0)}  .
  \label{RNCMetricExp}
\end{equation}
The coefficients here define tensor fields
\begin{equation}
  g_{ab,c_1 \cdots c_n}(x) = \left[  e^A_a e^B_b e^{C_n}_{c_n} \cdots e^{C_n}_{c_n}  \left( \frac{\partial^n \tilde{G}_{AB}}{\partial X^{C_1} \cdots \partial X^{C_n}} \right) \right]_{(x,0)}
  \label{Def-MetricExt}
\end{equation}
that are independent of the tetrad. They are referred to as metric normal tensors \cite{Dix74, Thomas}. The general process of looking at the tensor fields defined by coefficients in a Taylor series in Riemann normal coordinates is known as tensor extension. $g_{ab, c_1 \cdots c_n}$ is therefore the $n^{\mathrm{th}}$ tensor extension of $g_{ab}$.

Inverting \eqref{RNCMetricExp} with the help of \eqref{Def-RNC}, \eqref{Def-H}, \eqref{Def-RNCMetric}, and \eqref{Def-MetricExt}, the ordinary metric $g_{ab}(x)$ may be shown to have the expansion
\begin{equation}
  g_{a' b'} (x') \simeq \sigma^{a}{}_{a'} \sigma^{b}{}_{b'} \sum_{n=0}^N \frac{1}{n!}  X^{c_1} \cdots X^{c_n}  g_{a b,c_1 \cdots c_n}(x) .
  \label{MetricExpand}
\end{equation}
We have defined $X^a = - \sigma^a(x,x')$ by analogy to \eqref{Def-RNC}. The right-hand side of \eqref{MetricExpand} makes use of an origin $x$ that does not appear on the left. It represents the usual freedom in Taylor-expanding a function about an arbitrary point.

It is possible to deduce a number of general properties of the metric normal tensors. It is clear from their definition that
\begin{equation}
  g_{ab,c_1 \cdots c_n} = g_{(ab),c_1 \cdots c_n} = g_{ab,(c_1 \cdots c_n)}.
  \label{MetricExtSyms0}
\end{equation}
Less obviously, they also satisfy
\begin{equation}
  g_{a(b,c_1 \cdots c_n)} = g_{(a b,c_1 \cdots c_{n-1} ) c_n} = 0.
  \label{MetricExtSyms}
\end{equation}
In $p$ dimensions, the $n$th-order metric extension has
\begin{equation}
  \frac{p (n-1) (n+p-1)!}{2 (n+1)! (p-2)!}
\end{equation}
algebraically independent components \cite{Thomas, MetricExtComps}.

To prove \eqref{MetricExtSyms}, recall that the world function satisfies \cite{PoissonRev}
\begin{equation}
  \sigma^a \sigma_a = 2 \sigma.
  \label{SigIdent}
\end{equation}
Differentiating with respect to $x$ shows that $\sigma^{a'} \sigma^{a}{}_{a'} = \sigma^{a}$. Hence,
\begin{equation}
  H^{a'}{}_{a} \sigma^{a} = - \sigma^{a'} .
\end{equation}
Combining this with \eqref{Def-RNCMetric},
\begin{equation}
  X^A(x,x') [ G_{AB}(x,x') - \eta_{AB} ] = 0.
  \label{FockSchwinger}
\end{equation}
$\eta_{AB} = \tilde{G}_{AB}(x,0) = G_{AB}(x,x)$ is the ordinary Minkowski metric. Repeatedly differentiating \eqref{FockSchwinger} and evaluating the result at the origin recovers \eqref{MetricExtSyms}. Incidentally, this equation also provides an interesting link between Riemann normal coordinates and the Fock-Schwinger gauge of electrodynamics \cite{RNCFockSchwinger1, RNCFockSchwinger2}.

First derivatives of the metric vanish in Riemann normal coordinates, so $g_{ab,c}=0$. More generally, Dixon has found all metric normal tensors to linear order in the curvature \cite{Dix74}. For all $n \geq 2$,
\begin{equation}
  g_{ab,c_1 \cdots c_n} \simeq 2 \left( \frac{n-1}{n+1} \right) \nabla_{(c_3 \cdots c_n} R_{|a|c_1 c_2) b} + O (R^2) .
  \label{MetricExtLinear}
\end{equation}
This equation is exact if $n=2$ or $3$. Nonlinear terms appear at higher orders, however. It is extremely tedious to compute them by hand, although several methods have been developed to make the process conceptually straightforward \cite{RNCFockSchwinger1, RNCArxiv}. For $n=4$,
\begin{equation}
  g_{ab,c_1 c_2 c_3 c_4} = \frac{6}{5} \nabla_{(c_1 c_2} R_{|a|c_3 c_4)b} + \frac{16}{15} R_{a(c_1 c_2}{}^{d} R_{| b | c_3 c_4)d} .
\end{equation}
Going to much higher orders than this is best done with a computer algebra package \cite{RNCCadabra}.

In general, $g_{ab,c_1 \cdots c_n}$ can always be written as a polynomial in the metric, the Riemann tensor, and its first $n-2$ covariant derivatives (if $n \geq 2$). As indicated by \eqref{MetricExtLinear}, the linear term always involves exactly $n-2$ covariant derivatives of $R_{abc}{}^{d}$. Its remainder can include pieces with at least two fewer derivatives than this. These results may be inverted. \textit{Any number of covariant derivatives of the Riemann tensor may be written as a finite polynomial of metric normal tensors} (including the ``$n=0$ extension'' $g_{ab}$ and its inverse). This is easily seen by noting that $\nabla_{f_1 \cdots f_n} R_{abc}{}^{d}$ can always be written in terms of partial derivatives of the metric in an arbitrary coordinate system. Specifically, this makes sense in a Riemann normal coordinate system. But each of these partial derivatives may be replaced by a metric normal tensor on account of \eqref{Def-RNCMetric} and \eqref{Def-MetricExt}. This means that almost any covariant quantity locally constructed from the metric may be written entirely in terms of its extensions. This result is essentially Thomas' replacement theorem \cite{Thomas}.

\ack

I am grateful for many helpful comments and discussions with Robert
Wald and Samuel Gralla. This work was supported by NSF grant
PHY04-56619 to the University of Chicago.

\end{document}